\documentclass[%
 reprint,%
 amssymb, amsmath,%
 pre,cha,%
]{revtex4-1}

\usepackage{docs}%
\usepackage{bm}%
\usepackage[colorlinks=true,linkcolor=blue]{hyperref}%
\usepackage{graphicx}
\expandafter\ifx\csname package@font\endcsname\relax\else
 \expandafter\expandafter
 \expandafter\usepackage
 \expandafter\expandafter
 \expandafter{\csname package@font\endcsname}%
\fi
\usepackage{lipsum}

\usepackage{graphicx}
\usepackage{dcolumn}
\usepackage{bm}

\newcommand{\bbm}{\begin{multline}}
\newcommand{\eem}{\end{multline}}
\newcommand{\be}{\begin{equation}}
\newcommand{\ee}{\end{equation}}
\newcommand{\bea}{\begin{eqnarray}}
\newcommand{\eea}{\end{eqnarray}}

\newcommand{\comment}[1]{}

\newcommand{\xv}{\bm{{\rm x}}}

\newcommand{\vv}{\bm{{\rm v}}}

\newcommand{\qv}{{\bm{{\rm q}}}}

\begin{document}


\title{Hydrodynamic correlation functions of chiral active fluids}

\author{Debarghya Banerjee}
\email{debarghya.banerjee@ds.mpg.de}
\affiliation {Max Planck Institute for Dynamics and Self-Organization, 37077 G\"ottingen, Germany}
\altaffiliation{Instituut-Lorentz, Universiteit Leiden, Leiden 2300 RA, The Netherlands}
\author{Anton Souslov}%
\email{a.souslov@bath.ac.uk}
\affiliation{Department of Physics, University of Bath, Bath, UK
}%
\altaffiliation{James Franck Institute, The University of Chicago, Chicago, IL, USA}
\altaffiliation{Instituut-Lorentz, Universiteit Leiden, Leiden 2300 RA, The Netherlands}
\author{Vincenzo Vitelli}
\email{vitelli@uchicago.edu}
\affiliation{James Franck Institute, The University of Chicago, Chicago, IL, USA}
\altaffiliation[Also at ]{Department of Physics, The University of Chicago, Chicago, IL, USA}
\altaffiliation{Instituut-Lorentz, Universiteit Leiden, Leiden 2300 RA, The Netherlands}




\date{\today}

\begin{abstract}
The success of spectroscopy to characterise equilibrium fluids, for example the heat capacity ratio,
suggests a parallel approach for active fluids.
Here, we start from a hydrodynamic description of chiral active fluids composed of spinning constituents and
derive their low-frequency, long-wavelength response functions using the Kadanoff-Martin formalism. We find that the presence of odd (equivalently, Hall) viscosity leads to mixed density-vorticity response even at linear order. Such response, prohibited in time-reversal-invariant fluids, is a large-scale manifestation of the microscopic breaking of time-reversal symmetry. Our work suggests possible experimental probes that can measure anomalous transport coefficients in active fluids through dynamic light scattering.
\end{abstract}

\maketitle


\section{\label{sec:sec1} Introduction}
Spectroscopy of a fluid involves measuring linear response using scattering probes in order to characterize macroscopic modes and microscopic constituents. For example, scattering by electromagnetic waves directly measures the density-density response, via a quantity called the dynamic structure factor (DSF). The large-frequency, large-wavevector parts of the DSF (i.e., the scattering function obtained using either neutron or X-ray scattering) measure the inter-molecular correlations and interactions on the smallest scales. On the other hand, scattering by visible or near-visible light can measure the low-frequency, low-wavevector properties of simple fluids---precisely the properties captured by the equations of fluid hydrodynamics. This subtle relationship between the hydrodynamics and DSF was first derived by Landau and Placzek~\cite{Landau1934} for simple fluids and explored in generality in Ref.~\cite{Kadanoff1963} (see also Ref.~\cite{Chaikin1995}).

The dynamic structure factor contains information about macroscopic thermodynamic quantities (e.g., specific heat and compressibility) as well as response coefficients (e.g., diffusivity). Inertial density waves (i.e., acoustics) are well characterized by a region of DSF called the Brillouin peak---the peak location captures wave dispersion, and the peak width and height capture wave attenuation. On the other hand, the purely dissipative thermal response is contained in the Rayleigh peak of the DSF. These two peaks allow for the measurement of the ratio of isobaric ($C_P$) to isochoric ($C_V$) specific heats (equivalently, ratio of isothermal to adiadatic compressibilities) via the ratio of the peak heights, also called the Landau-Placzek ratio~\cite{Landau1934}.

The success of correlations and response to characterize equilibrium fluids suggests a parallel approach for the hydrodynamics of active fluids~\cite{Geyer2018}. To implement this idea, we characterize how the anomalous coefficients of active-fluid hydrodynamics enter the fluid's response. For example, chiral active fluids possess an anomalous transport coefficient called \emph{odd viscosity}~\cite{Banerjee2017,Souslov2019}. Such active fluids are composed of self-rotating particles, with examples including biological~\cite{Sumino2012,Tabe2003,Drescher2009,Petroff2015,Riedel2005}, colloidal~\cite{Snezhko2016,Maggi2015,Lemaire2008,Kokot2017}, granular~\cite{Tsai2005}, polymer~\cite{Denk2016}, and liquid-crystalline~\cite{Oswald2015} constituents. For isotropic spheres or disks, it may be difficult to measure single-particle rotations, but anomalous hydrodynamic coefficients can nevertheless reveal the active nature of fluid mechanics. These coarse-grained coefficients are present due to the effect of active rotations on the large-scale motion of the particles.
 Recent experimental advances have led to measurements of odd (Hall) viscosity in graphene's electron fluid~\cite{Berdyugin2019} and in chiral active fluids composed of spinning colloids~\cite{Soni2019,Abanov2019}.
In addition to odd viscosity~\cite{Banerjee2017,Avron1998}, such anomalous coefficients can include an anti-symmetric component to the fluid stress~\cite{Dahler1961,Condiff1964}.

How can one use scattering to distill the effects of odd viscosity from those of other viscosity coefficients and anti-symmetric stress? We answer this question using an analysis that parallels Refs.~\cite{Kadanoff1963,Chaikin1995}, but for chiral active hydrodynamics. Significantly, we find that odd viscosity leads to an anomalous dynamic response of vorticity due to density excitations and vice versa. This anomalous density-vorticity correlation distinguishes the effects of odd viscosity not only from equilibrium hydrodynamic coefficients, but also from the effects of anti-symmetric stress present in chiral active fluids. 

\section{\label{sec:sec2}HYDRODYNAMIC EQUATIONS OF ACTIVE ROTOR FLUIDS}

The emergent physics in systems of active rotors has recently been explored  using a variety of theoretical and numerical techniques~\cite{Lenz2003,Uchida2010,Yeo2015,Spellings2015,Nguyen2014,vanZuiden2016,Bonthuis2009,Furthauer2012}. The presence of torques in such chiral active fluids distinguishes these systems from the more commonly studied class of active materials: those composed of (polar) self-propelled particles. The presence of activity breaks time-reversal symmetry, whereas the presence of active rotation breaks parity in two-dimensional systems~\cite{Banerjee2017}. This breaking of symmetries leads to the breakdown of Onsager reciprocal relations that restrict fluid response. Specifically, the presence of anti-symmetric stress~\cite{Dahler1961,Condiff1964} and odd viscosity~\cite{Banerjee2017,Avron1998} in the hydrodynamic limit distinguishes active-rotor fluids from their polar active counterparts.

The presence of active rotation makes the system of chiral active rotors similar to a two-dimensional quantum system of charges in a magnetic field. As shown in Ref.~\cite{Banerjee2017}, one can find an emergent odd viscosity in these systems analogous to the Hall viscosity~\cite{Avron1995,Read2009} predicted in electronic quantum Hall fluids~\cite{Stern2008} and measured in graphene~\cite{Berdyugin2019}. The addition of Hall viscosity to hydrodynamic stress~\cite{LandauVI,Avron1998} results in the phenomenology discussed in Refs.~\cite{Wiegmann2014,Avron1998,Lapa2014,Ganeshan2017,Lucas2014,Lingam2014}. We examine the presence of odd viscosity and anti-symmetric stresses in chiral active fluids in which these terms emerge as a consequence of the coupling between intrinsic angular momentum and fluid velocity. Both odd viscosity and anti-symmetric stress show up in the transverse response of the fluid. However, these effects can be distinguished by the fact that odd viscosity depends on the \emph{mean} intrinsic rotation rate whereas hydrodynamic terms due to anti-symmetric stress only enter in proportion to \emph{spatial gradients} of the intrinsic rotation rate.

\begin{figure}
    \centering
    \includegraphics{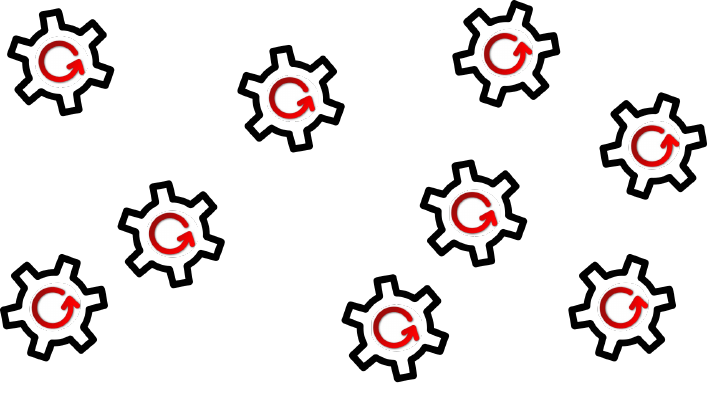}
    \caption{A schematic representation of a chiral active fluid composed of granular rotors. The red arrows indicate the intrinsic rotation field for each of the fluid's constituents around their own center of mass. The frictional coupling between rotors is represented by a gear-like shape.}
    \label{fig:schematic}
\end{figure}

In two dimensions, the hydrodynamic equations of chiral active rotors~\cite{Banerjee2017} governing the evolution of the slowly varying fields, namely mass density $\rho$, momentum density $\rho \vv$, and intrinsic angular momentum density $I \Omega$ are given by:
\begin{align}
D_t \rho &= - \rho \nabla \cdot \vv, \label{eq:rho} \\
\rho D_t v_i &= \partial_j \left( \sigma_{ij}^a + \sigma_{ij}^s \right) -\Gamma^v v_i, \label{eq:mom} \\
I D_t \Omega &= \tau + D^{\Omega} \nabla^2 \Omega -\Gamma^{\Omega} \Omega -\epsilon_{ij} \sigma_{ij}^a, \label{eq:om}
\end{align}
where, $D_t \equiv \partial_t + v_k \partial_k$ is the convective derivative. Equation~(\ref{eq:rho}) arises from the conservation of mass in the flow whereas Eq.~(\ref{eq:mom}) arises from the combination of linear momentum conservation and friction, where $\vv$ is the velocity, $\Gamma^v$ is a friction term that dissipates linear momentum, and $\sigma_{ij} \equiv \left( \sigma_{ij}^s + \sigma_{ij}^a \right)$ is the hydrodynamic stress term written in terms
of the symmetric part $\sigma_{ij}^s$ and the antisymmetric part $\sigma_{ij}^a$. In two dimensions, the antisymmetric part of a two-component tensor is proportional to the Levi-Civita symbol $\epsilon_{ij} = - \epsilon_{ji}$, with $\epsilon_{xy} = 1$. Equation~(\ref{eq:om}) describes the evolution of intrinsic angular momentum of the particles constituting the fluid. This angular momentum is not conserved and can be acquired from an external torque, converted to fluid vorticity, or dissipated by friction.  Here, $\Gamma^{\Omega}$ is the rotational friction, $D^{\Omega}$ is the rotational diffusion, and $\tau$ is the active torque. In the above equations, the components of the hydrodynamic stress $\sigma_{ij}$ can be written as:
\begin{align}
\sigma^a_{ij} &\equiv \frac{\Gamma}{2} \epsilon_{ij} (\Omega - \omega/2) \nonumber \\
\sigma^s_{ij} &\equiv -p \delta_{ij} + \eta_{ijkl} v_{kl} + \eta^o_{ijkl} v_{kl} \label{eq:str-s},
\end{align}
where $p$ is the hydrostatic pressure, $v_{kl} \equiv \left( \partial_l v_k + \partial_k v_l\right)/2$ is the strain-rate tensor, and the odd viscosity tensor $\eta^o_{ijkl}$ is given by the psuedo-scalar $\eta^o$ that we derive below~\cite{Banerjee2017}. The vorticity of the flow is given by $\omega = \nabla^* \cdot \vv \equiv \epsilon_{ij} \partial_i v_j$.

\section{\label{sec:sec3}Relation between local rotations and odd viscosity}

In this section, we develop a variational approach for the derivation of odd viscosity from dissipative coefficients. To this end, we begin with an energy functional analogous to the
Rayleigh dissipation function, but which includes coupling of intrinsic rotation
to flow velocity. Note that in the previous section the coupling term between Eq.~(\ref{eq:mom}) and
Eq.~(\ref{eq:om}) can be generated using an energy functional which has the form:
\begin{equation}
F_0 = \frac{\Gamma}{2} \left( \Omega - \omega/2 \right)^2.
\label{eq:f-inc}
\end{equation}

The simplest way to augment this functional such that it includes coupling between linear momentum and intrinsic rotation is: 
\begin{equation}
F = \frac{\Gamma}{2} \left[\Omega - \omega/2 + \alpha \nabla \cdot (\Omega \vv) \right]^2,
\label{eq:f}
\end{equation}  
where $- \alpha \nabla \cdot (\Omega \vv) \equiv \omega_{\rm ind}/2$ is an induced vorticity.
Using the product rule, the above expression becomes
\begin{equation}
F = \frac{\Gamma}{2}\left(\Omega - \omega/2 + \alpha \Omega \nabla \cdot \vv + \alpha (\vv \cdot \nabla ) \Omega \right)^2.
\end{equation}
Substituting this expression into the equation describing dynamics of the local rotation field and evaluating the Euler-Lagrange equation, we obtain
\begin{align}
\rho {\cal D}_t \Omega = &- \frac{\delta F}{ \delta \Omega} =  - \frac{\partial F}{ \partial \Omega} + \nabla_i \frac{\partial F}{\partial (\nabla_i \Omega)} 
\\
= &- \Gamma \left[\Omega - \omega/2 + \alpha \Omega \nabla \cdot \vv  \right] \nonumber \\
&+ \alpha \Gamma (\vv \cdot \nabla) \left[ \omega/2 - \alpha \Omega \nabla \cdot \vv - \alpha (\vv \cdot \nabla ) \Omega \label{eq:om-l} \right].
\end{align}
Note
that in the final expression in Eq.~(\ref{eq:om-l}), the first term is identical to the last term in
Eq.~(\ref{eq:om}), whereas the second term is higher order in either the
gradients of $\vv$ or in $\alpha$.  Therefore, Eq.~(\ref{eq:om}) appropriately
describes the large-scale dynamics of the local rotation field.

Similarly, we derive the dynamics of the velocity field, 
\begin{align}
{\cal D}_t v_i =& - \frac{\delta F}{ \delta v_i} =  -  \frac{\partial F}{ \partial v_i } +  \nabla_j \frac{\partial F}{ \partial (\nabla_j v_i) } \nonumber\\
\approx& \frac{\Gamma}{2} \nabla^*_i \left[ \Omega - \omega/2 + \alpha  \nabla \cdot (\Omega \vv)\right] + \alpha \Gamma \Omega \nabla_i \left[ \Omega - \omega/2 \right],  \label{eq:v-der}
\end{align}
where we discard all
terms of order $\alpha^2 \Omega^2$ to get to Eq.~(\ref{eq:v-der})---these
terms only contribute as corrections to the existing terms in the stress.  Note that whereas the first term in Eq.~(\ref{eq:v-der}) provides the expected correction to the \emph{antisymmetric} component of the stress, the second term in this equation couples local rotation to the flow velocity within the \emph{symmetric} component of the stress. To show that this equation includes a contribution from odd (or Hall) viscosity, we rearrange Eq.~(\ref{eq:v-der}) to find the expression
\begin{align}
\rho {\cal D}_t v_i = & \frac{\Gamma}{2} \nabla^*_i \left[ \Omega - \omega/2 + \alpha (  \vv  \cdot\nabla ) \Omega \right]
+ \frac{\alpha \Gamma}{2} (\nabla^*_i \Omega) (\nabla \cdot \vv)  \nonumber \\
& - \alpha \Gamma \Omega \nabla_i\Omega  - \frac{1}{2} \alpha \Gamma \Omega \left[ \nabla \omega + \nabla^* (\nabla \cdot \vv) \right], \label{eq:v-f}
\end{align}
where the first term is the new antisymmetric stress term, the second term couples compressible flow to the gradients in the local rotation, and third term may be rewritten in terms of $\nabla (\Omega^2)$ and, therefore, contributes to the symmetric stress $\sigma^s$. Significantly, the last term may be reexpressed using the two-dimensional identity $- \nabla^2 \vv^* = \nabla (\nabla^* \cdot \vv) + \nabla^* (\nabla \cdot \vv)$ as $\frac{1}{2} \alpha \Gamma \Omega \nabla^2 \vv^*$. Comparing this term to the odd viscosity contribution to the flow, $\eta^o \nabla^2 \vv^*$, we conclude that in the active-rotor liquid, the effective odd viscosity can be written in terms of the local rotation field and the coupling parameters as 
\begin{equation}
\eta^o = \frac{1}{2} \alpha \Gamma \Omega. 
\end{equation}
This equality relates the dissipationless odd viscosity $\eta^o$ to the dissipative coefficient $\Gamma$ of anti-symmetric stress.

Although the precise form of Eq.~(\ref{eq:v-f}) depends on the form for the effective free energy in Eq.~(\ref{eq:f}), odd viscosity arises only as a consequence of the single cross-term in $F$ of the form $\alpha \Gamma \Omega \omega (\nabla \cdot \vv)$. There are many forms of $F$ that can generate this cross-term, including alternative ways of writing down a complete square. For example, if we instead had taken 
\begin{equation}
F_{2} = \frac{\Gamma}{2}\left[\Omega - \omega - \alpha_2 \nabla \cdot (\omega \vv)\right]^2,
\label{eq:f2}
\end{equation}
then the cross-term $\alpha_2 \Gamma \Omega \omega (\nabla \cdot \vv)$ appears, leading to a similar expression for $\eta^o$ in terms of $\alpha_2$.

\subsection{\label{sec:sec4}Density-dependent coefficients}

In this subsection, we present an additional way to derive 
odd viscosity. Starting from the energy functional $F = \frac{\Gamma}{2} (\Omega - \omega)^2$, we consider the functional dependence of $\Gamma$ on density $\rho$. Next we consider slow variations of density in time and expand $\Gamma(\rho(t))$. We obtain
\begin{align}
\Gamma(\rho) &= \Gamma_0 + \Gamma_1 (\partial_t \rho) + \ldots, \nonumber \\
	     &= \Gamma_0 - \Gamma_1 (\nabla \cdot (\rho \vv)).
\end{align}
The term in the energy functional of the form $\Gamma_1 \Omega \nabla \cdot (\rho \vv)$ can be reexpressed as
\begin{align}
& \Gamma_1 \Omega \nabla \cdot (\rho \vv) \nonumber \\
& = \Gamma_1 \rho \nabla \cdot (\Omega \vv) + \Gamma_1 \Omega (\vv \cdot \nabla) \rho - \Gamma_1 \rho (\vv \cdot \nabla) \Omega.
\end{align}
The first term in the last line has been shown in the previous section to result in odd viscosity. The relevant term in the expression for $F$ has the form $\Gamma_1 \rho \nabla \cdot (\Omega \vv )$ or, equivalently, the form of the $\alpha$ term in Eq.~(\ref{eq:f}), with $\alpha = \Gamma_1 \rho/\Gamma_0$.

From the form of the stress in Eq.~(\ref{eq:str-s}), we focus on two contributions that distinguish active-rotor liquids from those that are well described by the Navier-Stokes equations: (1) the antisymmetric stress $\sigma^a$ that corresponds to a local torque on the center-of-mass motion of the rotors and (2) the odd viscosity $\eta^o$ that results from the breaking of time-reversal symmetry. These expressions allow us to establish the conditions in which the odd viscosity dominates over the antisymmetric stress~\cite{Banerjee2017}.  We can estimate both $\sigma^a$ and $\eta^o$ in terms of the angular frequency $\Omega_0 \equiv \tau/\Gamma^\Omega$ corresponding to the average of the local rotation field. Then, $\sigma^a \sim \Gamma \Omega_0$ and $\eta^o \sim \alpha \Gamma \Omega_0/2$ scale similarly with the applied torque $\tau$. However, note that only gradients of $\sigma^a$ enter Eq.~(\ref{eq:mom}).  By contrast, $\eta^o$ enters as a factor multiplying a strain rate.  Therefore, in a liquid in which the gradients of $\Omega$ are much smaller than $\Omega_0$, the odd viscosity contribution will be of a lower order in hydrodynamic variables than the odd stress terms. In such a liquid, we can consider those phenomena associated with odd viscosity without considering the odd stress.

\section{Equations of motion}

To analyse the linear hydrodynamic response for chiral active fluids, we start out with a nonlinear set of equations, Eq.~(\ref{eq:rho}--\ref{eq:om}). In this section, we write out all of the terms explicitly and then linearize these equations around the state with constant density and no flow. We then relate these linearized equations of motion to the correlations and response functions at long timescales and at large lengthscales. 

The full nonlinear hydrodynamics (including contributions from odd viscosity and antisymmetric stress) is described by
\begin{align}
\partial_t \varrho + \nabla \cdot (\varrho \vv) & = 0 \label{eq:cont1} \\
\partial_t (\varrho \vv) + \nabla \cdot (\varrho \vv \vv) &= -c^2 \nabla \varrho -\Gamma^v \vv + \eta \nabla^2 \vv \nonumber \\ 
& + \eta^o \nabla^2 \vv^* + \frac{\Gamma^\prime}{2} \nabla^* (\Omega^\prime- \omega/2) \label{eq:v1} \\
\partial_t (I \Omega^\prime ) + \alpha \nabla \cdot (\vv I \Omega^\prime) &= \tau^\prime + D^{\Omega\prime} \nabla^2 \Omega^\prime \nonumber \\ 
& - \Gamma^{\Omega\prime} \Omega^\prime - \Gamma^\prime (\Omega^\prime -\omega/2) \label{eq:om1}
\end{align}
where $\varrho(\xv, t)$ is the fluid-density field, $\vv(\xv, t)$ is the velocity field, $c$ is the speed of sound in the fluid, $\alpha$ is a coefficient that measures how far the system is from Galileans invariance ($\alpha = 1$ is Galilean invariant), and, as before, $\Gamma^v$ is a coefficient of substrate friction, $\eta$ is the (dynamic) dissipative viscosity, $\eta^o$ is the odd viscosity, $\Gamma^\prime$ is the ``gear factor'' which enters as the coefficient of antisymmetric stress, $\Omega^\prime$ is the local rotation rate for particles in the fluid, $I$ is the moment of inertia of each particle, $\tau^\prime$ is the active torque that each particle experiences, $D^{\Omega\prime}$ is the diffusivity of intrinsic rotation, and $\Gamma^{\Omega\prime}$ is the coefficient of single-particle rotational friction. Here, we have introduced the prime symbol to distinguish these `dynamic' response coefficients from the `kinematic' coefficients per unit density or unit moment of inertia, introduced below. Here, we set $\Gamma^v = 0$ and $\alpha = 1$.

\subsection{Linearized equations of motion}

We now linearize Eqs.~(\ref{eq:cont1}-\ref{eq:om1}) around the state $(\varrho,\vv,\Omega^\prime) = (\rho_0 + \rho,0 + \vv,\Omega_0 + \Omega)$ in $\rho$, $\vv$, and $\Omega$, where $\Omega_0 \equiv \tau^\prime/(\Gamma^{\Omega\prime} + \Gamma^\prime)$. We find
\begin{align}
\partial_t \rho &= - \rho_0 \nabla \cdot \vv \label{eq:cont2} \\
\partial_t \vv &= - c^2 \nabla \rho / \rho_0 + \nu \nabla^2 \vv + \nu^o \nabla^2 \vv^* + \frac{\Gamma}{2} \nabla^* (\Omega - \omega/2) \label{eq:v2} \\
\partial_t \Omega &= D^{\Omega} \nabla^2 \Omega - \Gamma^{\Omega} \Omega - \Gamma^r (\Omega -\omega/2), \label{eq:om2}
\end{align}
where $\Gamma \equiv \Gamma^\prime/\rho_0$, $\nu$[$\equiv \eta/\rho_0$] ($\nu^o$[$\equiv \eta^o/\rho_0$]) is the kinematic dissipative (odd) viscosity, $\Gamma^r \equiv \Gamma^\prime/I$, $\Gamma^\Omega \equiv \Gamma^{\Omega\prime}/I$, and $D^\Omega \equiv D^{\Omega\prime}/I$.

Using Helmholtz decomposition, it is convenient to express $\vv$ in terms of longitudinal and transverse components, $\vv = \vv_\ell + \vv_t$, where $\nabla \times \vv_\ell = 0$ and $\nabla \cdot \vv_t = 0$. Then, the vorticity $\omega = \nabla \times \vv_t$ and the compression $\nabla \cdot \vv = \nabla \cdot \vv_\ell$ determine the flow up to a choice of inertial reference frame. Using this decomposition, we rewrite Eqs.~(\ref{eq:cont2}-\ref{eq:om2}),
\begin{align}
\partial_t \rho &= - \rho_0 \nabla \cdot \vv \label{eq:cont3} \\
\partial_t (\nabla \cdot \vv) &= - c^2 \nabla^2 \rho / \rho_0 + \nu \nabla^2 (\nabla \cdot \vv) + \nu^o \nabla^2 \omega\label{eq:vl3} \\
\partial_t \omega &= (\nu + \Gamma/4) \nabla^2 \omega - \nu^o \nabla^2 (\nabla \cdot \vv) - \frac{\Gamma}{2} \nabla^2 \Omega \label{eq:vt3} \\
\partial_t \Omega &= D^{\Omega} \nabla^2 \Omega - \Gamma^{\Omega} \Omega - \Gamma^r (\Omega -\omega/2). \label{eq:om3}
\end{align}

To further distinguish between the different terms, it is useful to combine Eqs.~(\ref{eq:cont3},\ref{eq:vl3}) into a single equation for the density $\rho$. In addition, we have the fields vorticity $\omega$ and instrinsic rotation $\Omega$ for a total of three hydrodynamic equations:
\begin{align}
\partial_t^2 \rho &= c^2 \nabla^2 \rho + \nu \nabla^2 (\partial_t \rho) - \rho_0 \nu^o \nabla^2 \omega, \label{eq:r4} \\
\partial_t \omega &= (\nu + \Gamma/4) \nabla^2 \omega - \frac{\nu^o}{\rho_0} \nabla^2 (\partial_t \rho) - \frac{\Gamma}{2} \nabla^2 \Omega, \label{eq:vt4} \\
\partial_t \Omega &= D^{\Omega} \nabla^2 \Omega - (\Gamma^{\Omega} + \Gamma^r) \Omega + \Gamma^r\omega/2 \label{eq:om4}.
\end{align}
Note that Eqs.~(\ref{eq:r4}-\ref{eq:om4}) highlight the main difference between the anomalous coupling
due to odd viscosity and antisymmetric stress.
Whereas odd viscosity couples the transverse velocity $\omega$ (i.e., the vorticity $\nabla \times \vv$) to
the density field $\rho$, the antisymmetric stress couples $\omega$ to the intrinsic rotation $\Omega$.

\section{From hydrodynamics to structure and response}

The hydrodynamic equations provide information about the response at large length- and time-scales. For the density field, information about this response is encoded in a different form in the (complex) response function $\rho(\qv,z)/\rho(\qv)$ (i.e., response in frequency $z$ due to an initial density configuration $\rho(\qv)$ in terms of the wavevector $\qv$), and in the dynamic structure factor $S(q,z)$ where $q = |\qv|$ is the wavenumber and $z$ is the angular frequency. In equilibrium, the fluctuation-dissipation theorem states that the response $z \Re[\rho(\qv,z)/\rho(\qv)]$ is proportional to the dynamic structure factor $S(q,z)$. Kadanoff and Martin~\cite{Kadanoff1963} showed how to derive such structure and response functions from (generalized) hydrodynamic equations in an equilibrium fluid.

Here we perform this analysis for a chiral active fluid, which does not obey the conditions of equilibrium and can therefore have additional response functions. For example, in equilibrium, the response $\omega(\qv,z)/\rho(\qv)$ (relating the transverse component of velocity to the density) is zero. We show that in a chiral active fluid, this response function is nonzero and proportional to odd viscosity. The response $\rho(\qv,z)/\omega(\qv)$ obeys the generalized Onsager relation $\omega(\qv,z)/\rho(\qv) \propto -  \rho(\qv,z)/\omega(\qv)$ appropriate for fluids with broken time-reversal symmetry. Furthermore, the intrinsic rotational response $\Omega(\qv,z)/\omega(\qv)$ is proportional to the antisymmetric stress, and the coupling $\rho(\qv,z)/\Omega(\qv)$ requires both odd viscosity and antisymmetric stress. In addition to these off-diagonal responses, chiral active fluids have signatures of activity in the usual diagonal response functions $\rho(\qv,z)/\rho(\qv)$, $\omega(\qv,z)/\omega(\qv)$, and $\Omega(\qv,z)/\Omega(\qv)$. We derive analytical expressions for various responses, in a variety of physical limits. Beforehand, we review the Kadanoff and Martin approach for the Navier-Stokes equations.

\subsection{Review: from Navier-Stokes equations to the dynamic structure factor}

Ref.~\cite{Kadanoff1963} analyses Eqs.~(\ref{eq:r4}-\ref{eq:om4}) for the case $\Gamma = \nu^o = 0$. In this case, these equations are identical to the linearized Navier-Stokes equations in the compressible regime,
\begin{align}
\partial_t^2 \rho &= c^2 \nabla^2 \rho + \nu \nabla^2 (\partial_t \rho) , \label{eq:r5} \\
\partial_t \omega &= \nu \nabla^2 \omega.  \label{eq:vt5}
\end{align}
(We have ignored the field $\Omega$ because it is not a hydrodynamic variable for an equilibrium fluid.)

Note that the equations for the density and the transverse velocity can be analyzed independently. To arrive at response functions, we take Fourier transforms in both space and time, keeping both the dynamical terms that depend on ($\qv, z$) and the terms stemming from initial conditions that depend on $\qv$ only. We first consider the simpler case of the vorticity, which obeys the diffusion equation. For the diffusion Eq.~(\ref{eq:vt5}), the right-hand side transforms to 
\begin{equation}
\int_0^\infty d t \int d\xv \, e^{-i \qv\cdot\xv + i z t} \left[ \nu \nabla^2 \omega(\xv,t) \right] = - \nu q^2 \omega(\qv,z)
\end{equation}
and the left-hand side transforms to 
\begin{equation}
\int_0^\infty d t \int d\xv \, e^{-i \qv\cdot\xv + i z t} \left[ \partial_t \omega(\xv,t) \right] = i z \omega(\qv,z) + \omega(\qv)
\end{equation}
where $\omega(\qv) = \int d\xv \, e^{-i \qv\cdot\xv}\omega(\xv,0)$ is the Fourier transform of the vorticity field at time $t = 0$. This last term arises due to integration by parts, and is essential in the analysis of the response. The combined equation then reads
\begin{equation}
(- i z + \nu q^2)\omega(\qv,z) = \omega(\qv).
\end{equation}
Comparing this expression with the definition of the response function: $\omega(\qv,z)/\omega(\qv)$, we obtain the expression $\omega(\qv,z)/\omega(\qv) = (i z + \nu q^2)^{-1}$. The fluctuation-dissipation theorem relates the vorticity-vorticity correlation function $S_{\omega,\omega} \equiv \langle \omega(\qv,z) \omega(-\qv,-z) \rangle$ to the real part of the response $\omega(\qv,z)/\omega(\qv)$ via $S_{\omega,\omega} = \chi_\omega z \Re [\omega(\qv,z)/\omega(\qv)]$. The proportionality coefficient $\chi_\omega$ is the thermodynamic static susceptibility of the vorticity due to an external torque density $\tau(\qv)$, $\chi_\omega = \omega(\qv)/\tau(\qv)$. These thermodynamic prefactors depend on the details of the system, and may be significantly affected by activity. We will only write out the correlations up to such prefactors. Therefore, 
\begin{equation}
S_{\omega,\omega} \propto \frac{\nu q^2 z}{z^2 + (\nu q^2)^2}. \label{eq:soeq}
\end{equation}
This result for the correlation function is the main conclusion of this analysis, and in equilibrium fluids it may be possible to measure it directly via scattering. However, it is more common to focus on measuring the density-density correlations  $S_{\rho,\rho} \equiv \langle \rho(\qv,z) \rho(-\qv,-z) \rangle = \langle \delta \varrho(\qv,z) \delta \varrho(-\qv,-z)\rangle$. This correlation can be obtained using the Fourier transform of Eq.~(\ref{eq:r5}),
\begin{equation}
[-z^2 + c^2 q^2 - i z \nu q^2] \rho(\qv,z) = [- i z + \nu q^2] \rho(\qv).
\end{equation}
Solving the above equation, we find that the complex response function is given by
\begin{equation}
\frac{\rho(\qv,z)}{\rho(\qv)} = \frac{- i z + \nu q^2}{-z^2 + c^2 q^2 - i z \nu q^2}.
\end{equation}
The real part of this response, and the corresponding density-density correlation function $S_{\rho,\rho}(\qv,z)$ can be read off as
\begin{equation}
S_{\rho,\rho}(\qv,z) \propto z \Re \frac{\rho(\qv,z)}{\rho(\qv)} = \frac{c^2 q^4 \nu z}{(z \nu q^2)^2 + (z^2  - c^2 q^2)^2}.
\label{eq:sreq}
\end{equation}
This is one of the terms in the expression first derived by Landau and Placzek, corresponding to adiabatic sound propagation. This term dominates away from the critical point. The other part of the dynamic structure factor corresponds to heat transport and results in corrections to this expression near $q = 0$.
\begin{widetext}

\subsection{Chiral active fluids}
Now we extend the above analysis to chiral active fluids in a parallel approach. We start with the linearized equations of motion, Eqs.~(\ref{eq:r4}-\ref{eq:om4}), whose Fourier transforms are:
\begin{align}
[-z^2 + c^2 q^2 - i z \nu q^2] \rho(\qv,z) & = 
 [- i z + \nu q^2] \rho(\qv) + \nu^o \rho_0 q^2 \omega(\qv,z)\\
[- i z + (\nu + \Gamma/4) q^2]\omega(\qv,z) & = 
 \omega(\qv) - i z \nu^o q^2 \rho(\qv,z)/\rho_0 - \nu^o q^2 \rho(\qv)/\rho_0 + \frac{1}{2}\Gamma q^2 \Omega(\qv,z) \\
[- i z + \Gamma^{\Omega} + \Gamma^r + D^{\Omega} q^2]\Omega(\qv,z)&  = \Omega(\qv) + \Gamma^r \omega(\qv,z)/2. \label{eq:om6}
\end{align}
The response functions that result from this set of equations are easiest to analyze in separate two limits: the limit in which the fluid is dominated by antisymmetric stress ($\nu^o \rightarrow 0$), considered in the next section, and the limit in which the fluid is dominated by odd viscosity ($\Gamma \rightarrow 0$), considered in the following section.

\subsection{Structure in a chiral active fluids dominated by antisymmetric stress}

In a chiral active fluid in which the rotation rate is slow, gradients of the intrinsic rotation rate $\Omega$ and the resulting antisymmetric stress dominate over the higher-order response that involves a product of $\Omega$ and strain rates $\partial_i v_j$. As shown in the previous sections, the odd viscosity is a linearised version of this cross-coupling between $\Omega$ and $\partial_i v_j$. In the limit of slow rotation rate, we can consider odd viscosity to be negligible, $\nu^o \rightarrow 0$, and focus on the effect of antisymmetric stress only. Because the antisymmetric stress does not enter the density-density correlation function, in this case the expression for the dynamic structure factor is the same as for an equilibrium fluid. The other response functions can be calculated from the two equations for the transverse velocity and the intrinsic rotation rate $\Omega$:
\begin{align}
[- i z + (\nu+\Gamma/4) q^2]\omega(\qv,z) &= \omega(\qv) +\frac{1}{2}\Gamma q^2 \Omega(\qv,z) \\
[- i z + \Gamma^{\Omega} + \Gamma^r + D^{\Omega} q^2]\Omega(\qv,z) &= \Omega(\qv) + \Gamma^r \omega(\qv,z)/2
\end{align}
In order to solve this linear system of equations, we represent it as a matrix equation and invert the matrix:
\begin{equation}
\begin{pmatrix}
\omega(\qv,z) \\
\Omega(\qv,z)
\end{pmatrix}  = 
P(q,z) \begin{pmatrix}
(- i z + \Gamma^{\Omega} + \Gamma^r + D^{\Omega} q^2) &- \Gamma q^2 \\
- \Gamma^r & - 2 i z + (2\nu+\Gamma/2) q^2
\end{pmatrix}
\begin{pmatrix} 
\omega(\qv) \\
\Omega(\qv)
\end{pmatrix},
\end{equation}
where the prefactor $P(q,z)$ is given by
\begin{equation}
P(q,z) \equiv \frac{1}{ - \Gamma \Gamma^r q^2/2 +2 (\Gamma^{\Omega} + \Gamma^r + D^{\Omega} q^2 - i z) ([\nu+\Gamma/4] q^2 - i z)}.
\end{equation}

The entries in the inverted matrix equation are exactly the hydrodynamic response functions. Assuming that the fluctuation-dissipation theorem holds, we then relate these response functions to dynamic correlations $S_{a,b}$,
where each of the entries $a$ and $b$ can be either the field $\rho$ or $\omega$.
The correlation function is the ensemble average of the product of these two fields in Fourier space. The resulting expressions are:
\begin{align}
S_{\omega,\omega}(\qv,z) \propto  z \Re \frac{\omega(\qv,z)}{\omega(\qv)} = 
\frac{- 4 z q^2 \Gamma \Gamma^r (\Gamma^{\Omega} + \Gamma^r + D^{\Omega} q^2) 
+ 4 z [\nu+\Gamma/4] \{ (\Gamma^{\Omega} + \Gamma^r + D^{\Omega} q^2)^2 + z^2\}}
{[\Gamma \Gamma^r q^2 +2 (\Gamma^{\Omega} + \Gamma^r + D^{\Omega} q^2)[\nu+\Gamma/4] q^2 - 2 z^2]^2+4 z^2  [\Gamma^{\Omega} + \Gamma^r + D^{\Omega} q^2 + [\nu+\Gamma/4] q^2]^2},
\end{align}
\begin{align}
S_{\Omega,\Omega}(\qv,z) \propto z \Re \frac{\Omega(\qv,z)}{\Omega(\qv)} = 
\frac{- 2 z \Gamma \Gamma^r q^4  [2\nu+\Gamma/2] + 4 z (\Gamma^{\Omega} + \Gamma^r + D^{\Omega} q^2) ([\nu+\Gamma/4]^2 q^4 + z^2)}{[- \Gamma \Gamma^r q^2 +2 (\Gamma^{\Omega} + \Gamma^r + D^{\Omega} q^2)[\nu+\Gamma/4] q^2 - 2 z^2]^2+4 z^2  [\Gamma^{\Omega} + \Gamma^r + D^{\Omega} q^2 + [\nu+\Gamma/4] q^2]^2},
\end{align}
\begin{align}
S_{\omega,\Omega}(\qv,z) \propto  z \Re \frac{\omega(\qv,z)}{\Omega(\qv)} = 
\frac{ - 2 z \Gamma q^4 (-\Gamma \Gamma^r + 2 [\nu+\Gamma/4] (\Gamma^{\Omega} + \Gamma^r + D^{\Omega} q^2)) - 4 \Gamma q^2 z^3}{[-\Gamma \Gamma^r q^2 +2 (\Gamma^{\Omega} + \Gamma^r + D^{\Omega} q^2)[\nu+\Gamma/4] q^2 - 2 z^2]^2+4 z^2  [\Gamma^{\Omega} + \Gamma^r + D^{\Omega} q^2 + [\nu+\Gamma/4] q^2]^2}.
\end{align}
Note the Onsager relation $S_{\omega,\Omega}(\qv,z) \propto \Gamma^r S_{\Omega,\omega}(\qv,z) / (\Gamma q^2) = \rho_0 S_{\Omega,\omega}(\qv,z) / (I q^2)$.

In the trivial limit $\Gamma \rightarrow 0$, we find that the expression for $S_{\omega,\omega}$ reduces to Eq.~(\ref{eq:soeq}) with corrections $O([\Gamma]^2)$. To lowest order in $\Gamma$, the signatures of the antisymmetric stress are the correlation function $S_{\omega,\Omega}(\qv,z)$ and the response function $z \Re \frac{\omega(\qv,z)}{\Omega(\qv)}$, which are both linear in $\Gamma$:
\begin{equation}
S_{\omega,\Omega}(\qv,z) \propto  z \Re \frac{\omega(\qv,z)}{\Omega(\qv)} \sim - \frac{8 \Gamma q^2 ([\Gamma^{\Omega} + D^{\Omega} q^2]\nu q^2 - z^2)}{(\nu^2 q^4 + 
   z^2) ([\Gamma^{\Omega}  + D^{\Omega} q^2]^2 + z^2)} + O(\Gamma^3). \label{eq:lowgam}
\end{equation}
For a scattering experiment for a chiral active fluid dominated by gradients in $\Omega$ and therefore by antisymmetric stress, measuring the characteristic shape of the response in Eq.~(\ref{eq:lowgam}) would quantify the anomalous response of this chiral active fluid.

\subsection{Structure in a chiral active fluids dominated by odd viscosity}

We show that the hydrodynamic responses allow one to differentiate between the phenomena associated with odd viscosity and antisymmetric stress. The case dominated by odd viscosity corresponds to the parameters $\Gamma = 0$ and $\nu^o \ne 0$. This limit occurs in fluids in which gradients of $\Omega$ (and the associated antisymmetric stress) are much smaller than the odd viscosity term proportional to both $\Omega$ (without gradients) and strain rates $\partial_i v_j$. In this section, we find the effects of odd viscosity on the response and correlations in the active fluid. In the absence of the coupling $\Gamma$, the equation for $\Omega$ decouples from the other equations and the signature of odd viscosity is in the remaining $2\times2$ system of equations for density and transverse velocity on which we now focus. In Fourier space, these two equations read:
\begin{align}
& [-z^2 + c^2 q^2 - i z \nu q^2] \rho(\qv,z) = [- i z + \nu q^2] \rho(\qv) + \nu^o \rho_0 q^2 \omega(\qv,z), \\
& [- i z + \nu q^2]\omega(\qv,z) = \omega(\qv) - i z \nu^o q^2 \rho(\qv,z)/\rho_0 - \nu^o q^2 \rho(\qv)/ \rho_0.
\end{align}

We proceed as before, by transforming these two equations into a single matrix equation and inverting the matrix. In matrix form, the above equations read:
\begin{equation}
\begin{pmatrix}
\rho(\qv,z) \\
\omega(\qv,z)
\end{pmatrix} = Q(q,z)
\begin{pmatrix}
i q^4 (\nu^2 - [\nu^o]^2) + 2 q^2 \nu z - i z^2 & i q^2 \nu^o \rho_0\\
- i c^2 q^4 \nu^o/\rho_0 &z q^2 \nu + i c^2 q^2 - i z^2
\end{pmatrix}
\begin{pmatrix}
\rho(\qv) \\
\omega(\qv)
\end{pmatrix},
\end{equation}
where the prefactor $Q(q,z)$ is defined via
\begin{equation}
Q(q,z) \equiv \frac{1}{c^2 q^2 (z + i q^2 \nu) - z [q^4 ( [\nu^o]^2 - \nu^2) + 2 i q^2 \nu z + z^2]}.
\end{equation}

Significantly, the form of the above matrix allows us to conclude that $\frac{\rho(\qv,z)}{\omega(\qv)} = - \frac{\rho_0^2}{c^2 q^2}\frac{\omega(\qv,z)}{\rho(\qv)}$. This is a generalization of Onsager reciprocity for the case in which the fluid has broken time-reversal symmetry and therefore time-reversal-odd correlations can exist. These correlations are related, up to a prefactor, by the time-reversal operation and therefore by a minus sign.

The fluid has therefore characteristic response functions for density-density, vorticity-vorticity and the off-diagonal vorticity-density. Assuming the fluctuation-dissipation theorem holds, the corresponding correlation functions are then given by:
\begin{equation}
S_{\rho,\rho} \propto z \Re \frac{\rho(\qv,z)}{\rho(\qv)} = \frac{c^2 q^4 \nu z [q^4 \{[\nu^o]^2 - \nu^2\} + z^2]}{z^2 (z^2 - c^2 q^2 - q^4\{[\nu^o]^2 - \nu^2\})^2 
+ \nu^2 q^4 [c^2 q^2 - 2 z^2]^2}.
\end{equation}
\begin{equation}
S_{\omega,\omega} \propto z \Re \frac{\omega(\qv,z)}{\omega(\qv)} =  \frac{q^2 \nu z [c^4 q^4 - 2 c^2 q^2 z^2 + z^2 q^4 \{[\nu^o]^2 - \nu^2\}] + z^4}{z^2 (z^2 - c^2 q^2 - q^4\{[\nu^o]^2 - \nu^2\})^2 
+ \nu^2 q^4 [c^2 q^2 - 2 z^2]^2}.
\end{equation}
\begin{equation}
S_{\rho,\omega} \propto z \Re \frac{\rho(\qv,z)}{\omega(\qv)} =  \frac{q^4 \nu \nu^o \rho_0 z [c^2 q^2 - 2 z^2]}{z^2 (z^2 - c^2 q^2 - q^4\{[\nu^o]^2 - \nu^2\})^2 
+ \nu^2 q^4 [c^2 q^2 - 2 z^2]^2}.
\end{equation}

\begin{figure}
    \centering
    \includegraphics{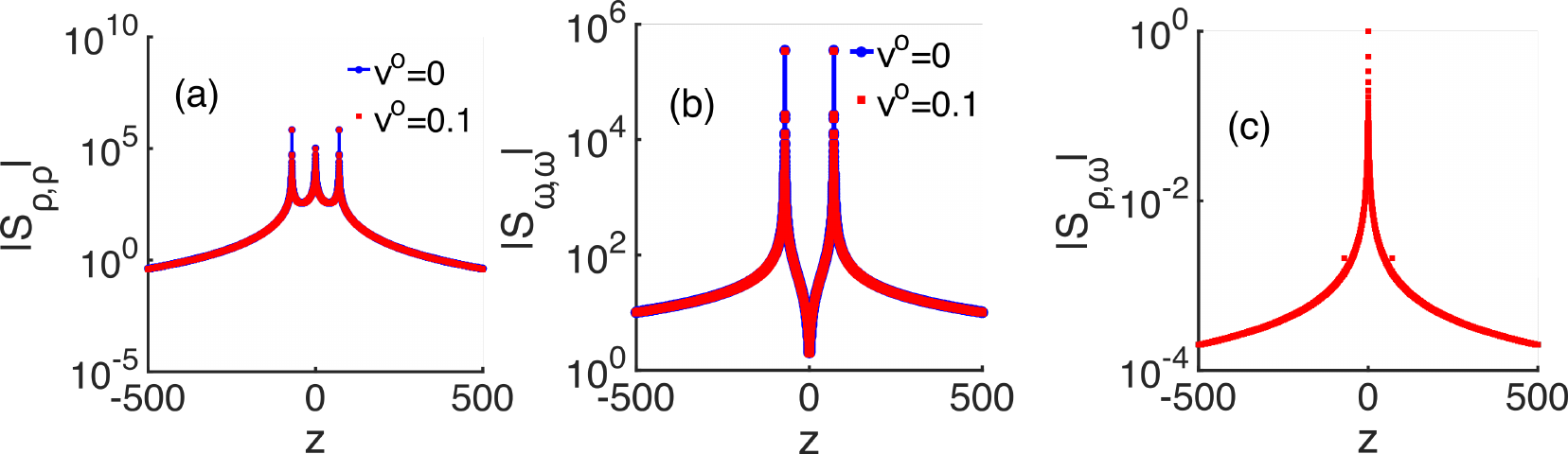}
    \caption{Correlation functions $S_{a,b}$ versus frequency of response $z$. (a) The density-density correlation functions for zero and small odd viscosity. (b) The vorticity-vorticity correlation function for zero and small odd viscosity. (c) The density-vorticity cross-correlation function, which has a nonzero value only in the presence of odd viscosity. Although in both (a) and (b) the effect of odd viscosity is negligibly small, this effect is apparent in part (c), because
    without odd viscosity this correlation function would be zero. In this figure, all of the parameters $\rho_0$, $\nu$, and $c^2$ are set to unity and we consider the wavenumber $q = 10$.}
    \label{fig:corr}
\end{figure}

In Fig.~\ref{fig:corr} we see the above mentioned correlation functions for $\nu^o = 0$ and small values of $\nu^o$. In this figure, all of the parameters  $\rho_0$, $\nu$, and $c^2$ are set to unity and we consider the wavenumber $q = 10$. 
This figure is described well by considering the case of small odd viscosity. In this case, $\nu^o \rightarrow 0$ and the expressions for $S_{\rho,\rho}$ and $S_{\omega,\omega}$ reduce to Eqs.~(\ref{eq:sreq}) and (\ref{eq:soeq}), respectively, with corrections $O([\nu^o]^2)$. To lowest order in odd viscosity, $O(\nu^o)$, the only effect of activity is the off-diagonal density-vorticity response and the density-vorticity correlation function $S_{\rho,\omega}$:
\begin{equation}
S_{\rho,\omega} \propto z \Re \frac{\rho(\qv,z)}{\omega(\qv)} \sim \frac{q^4 \nu \nu^o \rho_0 (c^2 q^2 - 2 z^2)}{(\nu^2 q^4  + z^2) [(z \nu q^2)^2 + (z^2  - c^2 q^2)^2]} + O([\nu^o]^3).  \label{eq:sros}
\end{equation}
This functional form is the main result of our work, showing the lowest-order change in fluid response and correlations due to the presence of odd viscosity.
This result suggests that a potential experiment to measure the dynamic correlation function in Eq.~(\ref{eq:sros}) could extract the value for odd viscosity.
\end{widetext}

In the case both odd viscosity and antisymmetric stress are present, the expressions become more complicated. Although the separate limits considered above capture most of the effect of odd viscosity and antisymmetric stress, there can be additional effects due to the combined effects of these two types of active stresses. In the case both $\Gamma \ne 0$ and $\nu^o \ne 0$, there exist nonzero correlations between the intrinsic rotation rate $\Omega$ and density $\rho$, $S_{\Omega,\rho}(\qv,z)$ (and the corresponding off-diagonal response function) proportional to $\nu^o \Gamma$.

\section{conclusions}

In this work, we have shown how anomalous linear transport coefficients arise in a fluid of spinning particles by expanding nonlinear anti-symmetric stress terms. We present several examples of this approach, using expansions in either the intrinsic rotation field or the density. These approaches all generate an anomalous transport coefficient called odd viscosity, which has been recently measured in both chiral active fluids~\cite{Soni2019} and electronic fluids in a magnetic field~\cite{Berdyugin2019}.

Starting from equations of hydrodynamics, we have derived the linearized response of chiral active fluids. We show that antisymmetric stress leads to off-diagonal response and correlations between the intrinsic spinning rate and vorticity. By contrast, the presence of odd viscosity leads to cross-correlations between \emph{density} and vorticity. This off-diagonal density-vorticity response results from the breaking of time-reversal symmetry and distinguishes odd viscosity from
other active hydrodynamic terms.  
Based on our results, we envision a general experimental approach for measuring odd viscosity using dynamical scattering of circularly polarized light.  In future work, these ideas could be extended to solids with odd elasticity~\cite{Scheibner2019}, viscoelastic fluids~\cite{Banerjee2020}, and anisotropic fluids with odd viscosity~\cite{SouslovAnton2019}.

\section{acknowledgements}
We gratefully acknowledge discussions with William Irvine and Tom Lubensky.
V.V.~was supported by the Complex Dynamics and
Systems Program of the Army Research Office under grant W911NF-19-1-0268. A.S. acknowledges the support of the Engineering and Physical Sciences Research Council (EPSRC) through New Investigator Award No. EP/T000961/1 and partial support through the Chicago MRSEC, funded by the NSF through grant DMR-1420709.


\end{document}